\documentclass[preprint,12pt]{elsarticle}

\usepackage{epsfig}

\usepackage{latexsym,amsmath,amscd,amssymb,graphics}
\usepackage{enumerate}

\usepackage{graphicx}
\usepackage{framed}
\usepackage[colorlinks]{hyperref}
\usepackage{url}
\usepackage{longtable}
\usepackage{nomencl}
\usepackage{amsmath}

\newcommand{\beq}{\begin{equation}} 

\newcommand{\eeq}{\end{equation}}

\def\be{\begin{equation}}
\def\ee{\end{equation}}
\def\bea{\begin{eqnarray}}

\usepackage{amssymb}

\journal{Elsevier}

\begin{document}

\begin{frontmatter}



\title{}

%-------------

\title{On the form of the power equation for modeling solar chimney power plant systems}

 \author[label1]{Nima Fathi\corref{cor1}}
\author[label2]{Seyed Sobhan Aleyasin}
\author[label1]{Peter Vorobieff}

 \address[label1]{The University of New Mexico}
 \address[label2]{The University of Manitoba}
 \cortext[cor1]{Corresponding author. Email: \texttt{nfathi@unm.edu}}

%\address{}

\begin{abstract}

Recently several mathematical models of a solar chimney power plant were derived, studied for a variety of boundary conditions, and compared against CFD calculations. The importance of these analyses is about the accuracy of the derived pressure drop and output power equation for solar chimney power plant systems (SCPPS). % by Koonsrisuk and Chitsomboon (Energy 51 (2013), 400-406). 
We examine the assumptions underlying the derivation and present reasons to believe that some of the derived equations, specifically the power equation in this model, may require a correction to be applicable in more realistic conditions. The analytical resutls are compared against the available experimental data from the Manzanares power plant.

%-----------------------

\end{abstract}


\end{frontmatter}

\section*{Nomenclature}
\begin{tabbing}

$Variables$\\
$A$ \hspace{0.25in}\= cross-sectional area\\
$A_r$ \>cross-sectional area of the collector ground \\
$g$ \> acceleration due to gravity \\
$h$ \> height\\
$\dot{m}$ \> air mass flow rate\\
$p$ \> pressure\\
$\dot{W}$ \> flow power \\
$q$ \> heat transfer per unit mass\\
$q''$ \> heat flux\\
$R$ \> gas constant \\
$T$ \> temperature\\
$\rho$ \>density\\
$u$ \> velocity\\
%$F_\mu$ \> friction force\\
$c_p$\>specific heat capacity\\
\\
$Subscripts$\\
$i$ \> inlet \\
$o$ \> outlet \\
$c$ \> collector \\
$t$ \> tower \\
%$cyl$\>cylindrical\\
$turb$\>turbine\\
$atm$\>atmospheric\\
\\
$Abbreviations$\\
$LHS$ \> \ left hand side\\
$RHS$ \>  \ right hand side\\

\\

\end{tabbing}



\section{Introduction}

\label{}

Although the idea of the solar chimney power plant (SCPP) can be traced to the early 20th century, 
practical investigations on solar power plant systems started in the late 1970s, around the time of
conception and construction of the first prototype in Manzanares, Spain. This solar power plant 
operated between 1982 and 1989 and the generated electric power was used in the local electric 
network \cite{1,2}. 

\begin{figure}  [!h]
\centerline{\includegraphics[width=4in, height=2.5in]{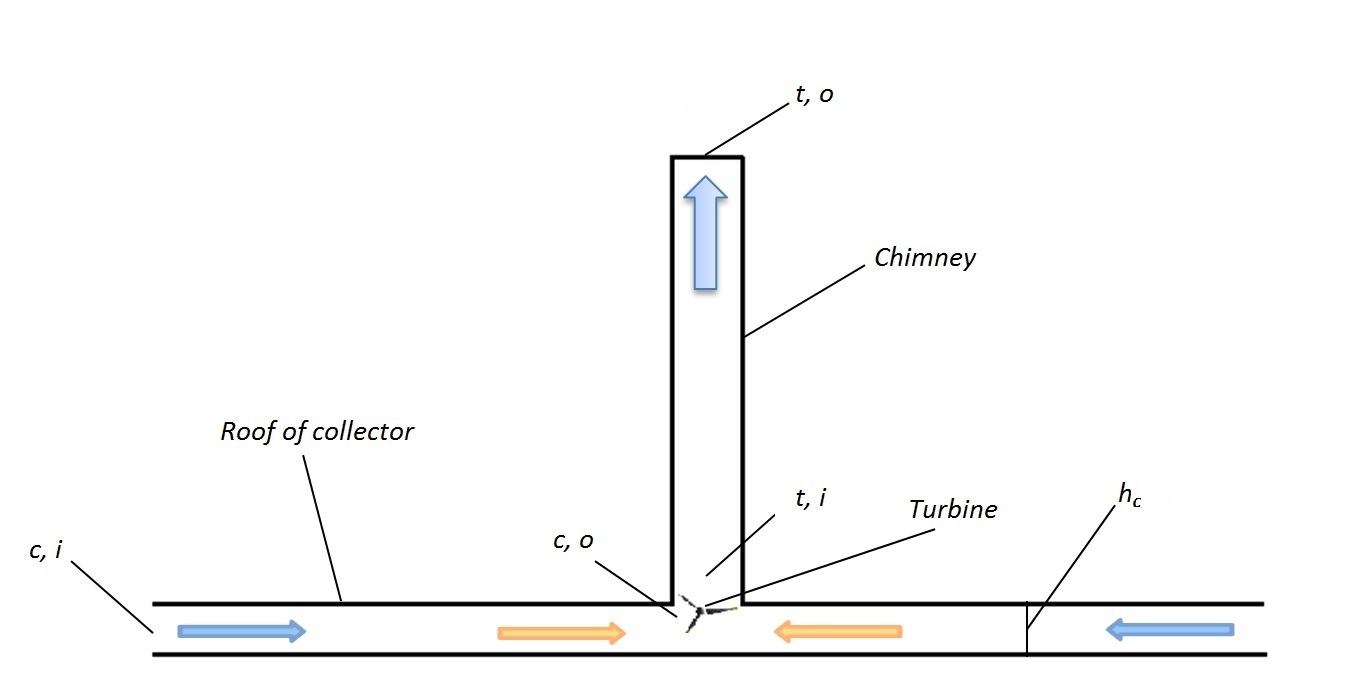}}
\caption{Schematic of SCPP with the applied variables and subscripts in the present analysis \label{fig1}}
\end{figure}

The basic SCPP concept (Fig.~\ref{fig1}) demonstrated in that facility is fairly striaghtforward. Sunshine heats the air beneath a transparent roofed collector structure surrounding the central base of a tall chimney tower. The hot air produces an updraft flow in the chimney. The energy of this updraft flow is harvested with a turbine in the chimney, producing electricity. 
Experiments with the prototype proved the concept to be viable, 
and provided data used by a variety of later researchers. A major motivation for subsequent studies lay in the need for reliable modeling of the operation of a large-scale power plant. The Manzanares prototype had a 200~m tall chimney and a 40,000~m$^2$ collector area. Proposals for economically competitive SCPP facilities usually feature chimneys on the scale of 1~km and collectors with multiple square kilometer areas.

Padki and Sherif \cite{3} used the results from the Manzanares prototype to extrapolate the data to large scale models for SCPP.  In 1991, Yan et al. \cite{4} developed an SCPP model using a practical correlation. They introduced equations including air velocity, air flow rate, output power, and thermofluid efficiency. Von Backstr\" om and Fluri conducted a numerical study to determine the optimum ratio of pressure drop of the turbine as a fraction of the available pressure difference required to achieve the maximum power \cite{5}. They noted that this ratio might lead to overestimating the flow passage in the plant and also designing a turbine without a sufficient stall margin. In other recent works, the SCPP concept involving an inflatable tower was examined, with all parts of the power plant modeled numerically \cite{8,9,10}. A small-scale inflatable tower was fabricated for validation of these results, and code calibration was performed using the newly available experimental data 
\cite{11,12}.

To find the maximum power, different atmospheric pressure and temperature boundary conditions were applied for various tower heights and atmospheric lapse rates \cite{13}. Theoretical analysis to study the effect of pressure drop in the  SCPP turbine was performed by Koonsrisuk et al. \cite{14}. The optimal pressure drop ratio was found numerically and analytically by Gue et al., around 0.9 for the Manzanares prototype. This inverstigation can be applied as an initial estimation for various SCPP turbines \cite{15}. Tayebi et al. modeled and simulated the SCPP with a curved conjunction between tower and collector for different Rayleigh numbers \cite{16}.

Earlier modeling efforts \cite{9} showed a keen sensitivity of the 
predictions
of SCPP output to boundary conditions, in particular, pressure. Numerical 
simulations require careful validation and verification, and for that, 
analytical models are indispensable. A theoretical model was recently 
developed \cite{kc13} to model the combined performance of the solar 
collector, chimney, and turbine. Here we will examine some of the 
assumptions and derivations in this model and present an alternative 
formulation for the energy equation. 

\section{Derivation of equations}

%\label{}

\subsection{Collector}

To derive the equations, we start from the collector. It is assumed that the flow through the collector 
is one-dimensional, steady-state, and compressible. Let us disregard the friction and assume the total heat from the solar irradiation is absorbed within the air filling the collector. For this one dimensional axisymmetric compressible flow analysis, the mass conservation satisfies:

\beq
 \frac{dA}{ A} + \frac{d\rho }{\rho }+\frac{du} u=0\quad \mbox{(Continuity)}
\label{cont-N}
\eeq

 Here $A$ is the cross-sectional area of the collector that air goes 
 through -- $A=2\pi r h_c$ and $dA=2\pi r dh_c$. 

Momentum equation is as follows \cite{a}:

%\color{green}

%\beq
% \frac { d(\rho u u)} {dr}+ \frac {dP}{dr}=0 %  \quad \mbox{(Momentum)}
%\label{mom1-N}
%\eeq

%\color{black}

The momentum equation can be obtained from:

\beq
 \frac {dp} {dr}+ \frac{\rho udu}{dr}=0 % \quad \mbox{(Momentum)}
\label{mom-N}
\eeq

\beq
 dp+\rho udu=0 \quad \mbox{(Momentum)}
\label{mom-N}
\eeq

\color{black}

The energy equation and the equation of states are

\beq
\frac {\rho u de}{dr} + \frac {pdu} {dr}+\frac {k d^2 T}{dx^2} =0
\label{Energy-N}
\eeq

\beq
  p=p(\rho,T) , e=e(\rho,T)
    \label{EnergyST-N}
\eeq

The last two equations,\eqref{EnergyST-N},  represent the thermal and caloric equations of state \cite{b}. Consider the energy balance equation and the equation of state as follows:

\beq
 c_pdT-dq+udu=0\quad \mbox{(Energy)}
\label{Energy-N}
\eeq

\beq
\frac{dp} p-\left(\frac{d\rho }{\rho }+\frac{dT} T\right)=0 \quad \mbox{(State)}
\eeq

To find $dp$ we can apply Eq.~\eqref{mom-N} and substitute 
${du}/{u}$ from the continuity equation, Eq.~\eqref{cont-N}.

\beq
-\frac {dp}{ \rho u^2}=\frac {du}{u}=-\frac{d\rho}{\rho}-\frac{dA}{A}
\label{dp-N1}
\eeq

\beq
{dp}={\rho u^2} \left(\frac{d\rho}{\rho}+\frac{dA}{A} \right)
\label{dp-N2}
\eeq

From the equation of state we can find $d\rho/\rho$ and substitute in Eq.~\eqref{dp-N2},

\beq
\frac {d\rho}{\rho}=\frac {dp}{p}-\frac {dT}{T}
\label{dp-N3}
\eeq

\beq
{dp}={\rho u^2} \left(\frac {dp}{p} -\frac {dT}{T}+ \frac{dA}{A} \right)
\label{dp-N4}
\eeq

We can rewrite Eq.~\eqref{dp-N4} as a function of $T, A, u, p, \rho$ , where $ \dot m=\rho A u$:

\beq
{dp}=\frac{\rho^2 u^2 A^2}{\rho} \left (\frac {dp}{A^2p} -\frac {dT}{A^2T}+ \frac{dA}{A^3} \right)
\label{dp-N5}
\eeq

Also by substitution $dT$ from the energy equation on the base of $dq, c_p$ and $u$,  we obtain

\beq
{dp}=\frac{\dot m^2}{\rho} \left( \frac{dA}{A^3}  -\frac {dq-udu}{A^2Tc_p}+\frac {dp}{A^2p}\right )
\label{dp-N6}
\eeq

For consistency with previous analyses, let us rewrite $dq$ on the basis 
of heat flux per mass flow rate---$dq=q'' dA_r/\dot m$ where $q$ has the units of $J/kg$. Here $A_r=\pi r^2$, therefore $dA_r=2\pi r dr$. Note that $A=2\pi r h_c $, where $h_c$ is the collector height (roof height) that was assumed to be proportional to r -- $h_c=ar$, where $a$ is a constant. By substituting $A_r$, $dq$ and $A$ in the second term on the RHS, we obtain

\beq
{dp}=\frac{\dot m^2}{\rho} \left(\frac{dA}{A^3} -\frac {q''(2\pi r)dr}{\dot m (2\pi r^2 a )^2Tc_p}+ \frac {udu}{A^2 c_pT}+\frac {dp}{A^2p} \right )
\label{dp-N8}
\eeq

We can rewrite equation \eqref{dp-N8} and substitute $udu$ of the third term on the RHS by applying momentum equation
 \eqref{mom-N}, $udu=-{dp}/{\rho}$.

\beq
{dp}=\frac{\dot m^2}{\rho} \left(\frac{dA}{A^3} -\frac {q''dr}{2 \pi \dot m   r^3 a^2  c_p T}- \frac {dp}{ A^2 \rho c_pT}+\frac {dp}{A^2p} \right )
\label{dp-N9}
\eeq

Then we can substitute $p$ from the equation of state, $p=\rho R T$ 
 and rewrite the above equation to find $dp$ on the LHS.

\beq
{dp}=\frac{\dot m^2}{\rho} \left[\frac{dA}{A^3} -\frac{q''dr}{2 \pi \dot m   r^3 a^2  c_p T}\right]{\left[1-  \frac{\dot m^2}{A^2 \rho^2 T}  \left(\frac1{R}-\frac1{c_p}\right)\right]}^{-1}
\label{dp-N10}
\eeq

Note that $$\dot m=\rho A u,$$ So we can rewrite $\eqref{dp-N10}$ as 

\beq
{dp}=\frac{\dot m^2}{\rho} \left[\frac{dA}{A^3} -\frac{q''dr}{2 \pi \dot m   r^3 a^2  c_p T}\right]{\left[1-  \frac{u^2}{ T}\left(\frac1{R}-\frac1{c_p}\right)\right]}^{-1}
\label{dp-N100}
\eeq

Equations \eqref{dp-N10} and  \eqref{dp-N100} are the exact 
solutions for $dp$ for the one-dimensional 
frictionless analysis of the collector. Since our fluid is air we can estimate $c_p$ and rewrite Eq.~$\eqref{dp-N100}$.

\beq
{dp}\simeq \frac{\dot m^2}{\rho} \left(\frac{dA}{A^3} -\frac {q''dr}{2 \pi \dot m   r^3 a^2  c_p T}\right){\left(1-\frac{2.494u^2}{
T}\right)}^{-1}
\label{dp-N110}
\eeq

On the basis of mass flow rate, Eq.~\eqref{dp-N110} can be written as:

\beq
{dp}\simeq \frac{\dot m^2}{\rho} \left(\frac{dA}{A^3} -\frac {q''dr}{2 \pi \dot m   r^3 a^2  c_p T}\right){\left(1-\frac{2.494{\dot m}^2}{
T\rho^2A^2}\right)}^{-1}
\label{dp-N11}
\eeq

The third term of the RHS of Eq.~$\eqref{dp-N11}$  was ignored \cite{kc13} which can be correct when density is constant. $c_p$, $q''$ and $T$ are considered approximately constant as well. Therefore by integrating between the inlet and outlet of the collector without the last term of the RHS, pressure difference can be derived. 

\beq
\int_{c,o}^{c,i} {dp} \simeq \int_{c,o}^{c,i} \left(\frac{\dot m^2dA}{\rho A^3} - \frac{\dot m q''dr}{2 \pi r^3 a^2  \rho c_p T}\right )
\label{dp-N12}
\eeq

\beq 
 p_{c,i}-p_{c,o}\simeq \left [ \frac{\dot m^2} {2\rho} \left(\frac 1{A_{c,o}^2}-\frac 1{A_{c,i}^2}\right)-\frac{q^{''}\dot m}{4\pi a^2\rho c_pT}\left(\frac 1{r_{c,o}^2}-\frac 1{r_{c,i}^2}\right)\right] 
 \label{P10}
\eeq

\subsection{Tower}

The air flow in the tower(chimney) is considered as an adiabatic frictionless flow. The conservation equations for the one-dimensional steady state flow in variable-area tower are as follows:

\beq
 \frac{d\rho }{\rho }+\frac{du} u+\frac{dA} A=0\quad \mbox{(Continuity)}
\eeq
\beq
 \frac{dP}{\rho }+gdz+udu=0 \quad \mbox{(Momentum)}
\eeq
\beq
 c_pdT+udu+gdz=0 \quad \mbox{(Energy)}
\label{Energy-NN}
\eeq
\beq
\frac{dp} p-\frac{d\rho }{\rho }-\frac{dT} T=0 \quad \mbox{(State)}
\eeq

By following the same trend to find $dp$ we get

\beq
{dp}=\left[-\rho g dz+ \frac {\dot m^2 dA}{\rho A^3}+ {\rho u^2} \left(\frac{dp}{p} -\frac{dT}{T}\right )\right]
\label{dp-T-N}
\eeq

By applying the energy equation and substitution $ dT={(-gdz-udu)} /{c_p} $, we can rewrite the above equation as

\beq
{dp}= \left[-\rho g dz+ \frac {\dot m^2 dA}{\rho A^3}+ {\rho u^2} \left(\frac{dp}{p} +\frac{gdz+udu}{c_pT}\right )\right]
\label{dp-T-N}
\eeq

Here $dp=-\rho(udu+gdz)$, then we get

\beq
{dp}= \left[-\rho g dz+ \frac {\dot m^2 dA}{\rho A^3}+ {\rho u^2} \left(\frac{dp}{p} -\frac{dp}{\rho c_pT}\right)\right]
\label{dp-T-N}
\eeq
%%%%%%%%%%%%%%%%%%%%%%%%%%%%%%%%%%%%%%%%%%%%%%%%%%%%%%%
Above equation can be solved for $dp$,

\beq
{dp}= \left[-\rho g dz+ \frac {\dot m^2 dA}{\rho A^3}\right]{\left[1-  \frac{u^2}{ T}\left(\frac1{R}-\frac1{c_p}\right)\right]}^{-1}
\label{dp-T-N}
\eeq

Also by considering the material properties of air the same way we did for 
the collector part,

\beq
{dp}\simeq \left[-\rho g dz+ \frac {\dot m^2 dA}{\rho A^3}\right]{\left[1-  \frac{2.494u^2}{ T}\right]}^{-1}
\label{dp-T-N}
\eeq
The above equation is the exact closed form solution of $dp$ at any point on the base of variable $\rho$, $T$ and $q''$.
We ignore the last term on the RHS by having a constant density \cite{kc13} and integrate between the inlet and outlet tower area to find the pressure difference of the chimney as,

\beq
\int_{t,0}^{t,i} {dp}\simeq \int_{t,0}^{t,i} \left( -\rho g dz+ \frac {\dot m^2 dA}{\rho A^3}\right)
\label{dp-T-N}
\eeq

\beq
p_{t,i}\simeq p_{t,o}+\rho gh_t+ \frac{\dot m^2}{2\rho} \left (\frac 1{A_{t,o}^2}-\frac 1{A_{t,i}^2}\right)
\label{P11}
\eeq

To calculate the output power, we can define the power on the basis of the pressure difference at the turbine -- where it is normally utilized at the outlet of the collector and inlet of the tower. 

\beq
\dot W \simeq \frac{\dot m(p_{c,o}-p_{t,i})}{\rho_{turb}} 
\label{w11}
\eeq

Let $\rho_{turb}=({\rho_{c,o}+\rho_{t,i}})/{2}$ and substitute equations $p_{c,o}$ and $p_{t,i}$ from \eqref{P10} and \eqref{P11}. Hence for the flow  power by assuming $p_{c,i}=p_{t,o}+\rho gh_t$, we have

\begin{eqnarray}
%\beq
\dot W= \frac {\dot m} {(\rho_{c,o}+\rho_{t,i})/2}  \Biggl[ \frac {-\dot m^2} {2\rho} \left (\frac1{A_{c,o}^2}- \frac 1{A_{c,i}^2} \right)  + \frac {q'' \dot m}{4 \pi a^2 \rho c_pT} \left (\frac {1}{r_{c,o}^2}-\frac {1}{r_{c,i}^2} \right) - \frac {\dot m^2}{2\rho} \left (\frac {1} {A_{t,o}^2}-\frac {1}{A_{t,i}^2}\right)\Biggr] 
  \label{w22}
%\eeq
\end{eqnarray}

For area the following equations are used, where $b$ and $c$ are arbitrary positive real constants.

\beq
 A^2_{c,i}=bA^2_{c,o},    \    A^2_{t,o}=cA^2_{t,i}
\label{a12}
\eeq

The simplified form of equation \eqref{w22} by applying the area correlations is,

\begin{eqnarray}
%\beq
\dot W \simeq \frac {\dot m} {( \rho_{c,o}+\rho_{t,i} ) /2 }\Biggl[\frac  {-\dot m^2}{2\rho} \left( \frac {b-1}{bA_{c,o}^2} + \frac {1-c}{cA_{t,i}^2} \right) + \frac{q^{''}\dot m}{4\pi a^2\rho c_pT}\left(\frac 1{r_{c,o}^2}-\frac 1{r_{c,i}^2}\right)   \Biggr]
 \label{w222}
\end{eqnarray}

%\eeq

%\begin{eqnarray}
%\beq
%\dot W \simeq \frac{\dot m}{(\rho_{c,o}+\rho_{t,i})/2} \Biggl[{-\frac{\dot m^2} {2\rho} \left(\frac {b-1}{bA_{c,o}^2} +\frac {1-c}{cA_{t,i}}\right) \nonumber 
%\\+\frac{q^{''}\dot m}{4\pi a^2\rho c_pT}\left(\frac 1{r_{c,o}^2}-\frac 1{r_{c,i}^2}\right) \right)\Biggr]}

% \label{w222}
%\end{eqnarray}

%\eeq

Assume $\rho$ is constant before and after turbine, therefore 

$$\rho_{turb}=\rho_{c,o}=\rho_{t,i}=\rho$$

Koonsrisuk et al. derived an equation in which the second term was neglected in comparison with the first term on the RHS of Eq.~\eqref{w222}. However, Eq.~\eqref{w14} shows the derived power equation by them at the end is likely to exceed the expected amount by a factor of two. 

\beq
\dot W\simeq \frac{-\dot m^3}{2\rho ^2}\left(\frac{1-c}{cA_{t,i}^2}+\frac{b-1}{bA_{c,o}^2}\right)
\label{w14}
\eeq

\begin{figure}  [!h]
\centerline{\includegraphics[width=4in, height=2.5in]{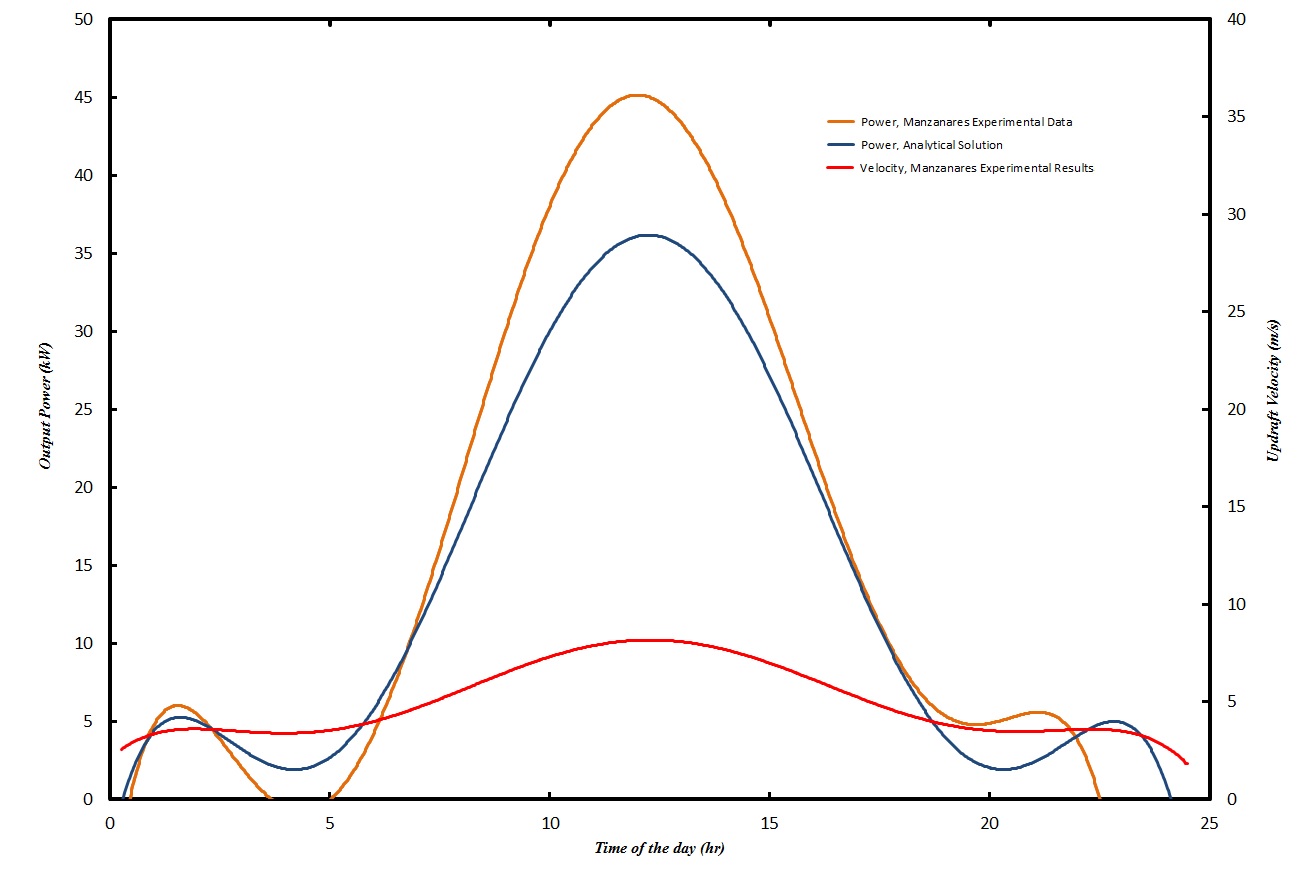}}
\caption{Analytical power results against measurement from Manzanares: updraft velocity and
power output for a typical day \label{fig2}}
\end{figure}

To evaluate the derived anlaytical solution for the output power of SCCP, the available experimental data from Manzanares prototype was applied and extracted. The measured updraft velocity for 25 hours Manzanares power plant operation is imposed to the analytical solution and the analytical power compared against the experimental outpower from the turbine (Fig.~\ref{fig2}).

\section{Conclusion}
We presented considerations supporting our contention that a
known analytical 
model predicting a solar chimney power plant performance may have a power equation which is off by a factor of two. Careful derivation of the models
is very important, especially for the specific area of interest related to
solar-chimney power plants, where numerical model scalability is a key issue,
and few experimental results are available for validation. 
During the verification and validation process, the modeler must ask two 
questions: "Am I modeling the physics correctly?" and "Am I modeling the
correct physics?" Comparison with analytical models is important for 
answering both of these questions, and the only way to have them well-posed 
is to have correct physics in the analytics.


\begin{thebibliography}{00}


\section*{References}

\bibitem{1}

Haaf, W., K. Friedrich, G. Mayr, and J. Schlaich. "Solar chimneys part I: principle and construction of the pilot plant in Manzanares." International Journal of Solar Energy 2, no. 1 (1983): 3-20.

\bibitem{2}

Haaf, W. "Solar chimneys: part ii: preliminary test results from the Manzanares pilot plant." International Journal of Sustainable Energy 2, no. 2 (1984): 141-161.

\bibitem{3}

Padki, M. M., and S. A. Sherif. "Solar chimney for medium-to-large scale power generation." In Proceedings of the manila international symposium on the development and management of energy resources, vol. 1, pp. 432-437. 1989.

\bibitem{4}

Yan, M. Q., S. A. Sherif, G. T. Kridli, S. S. Lee, and M. M. Padki. "Thermo-fluid analysis of solar chimneys." In Industrial Applications of Fluid Mechanics-1991. Proceedings of the 112th ASME winter annual meeting, Atlanta, GA, pp. 125-130. 1991.

\bibitem{5}

Von Backstr\"om, Theodor W., and Thomas P. Fluri. "Maximum fluid power condition in solar chimney power plants–an analytical approach." Solar Energy 80, no. 11 (2006): 1417-1423.




\bibitem{8}

Schlaich, Jörg, Michael Robinson, and Frederick W. Schubert. The solar chimney: electricity from the sun. Geislingen, Germany: Axel Menges, (1995).

\bibitem{9}

Putkaradze, Vakhtang, Peter Vorobieff, Andrea Mammoli, and Nima Fathi. "Inflatable free-standing flexible solar towers." Solar Energy 98 (2013): 85-98.


\bibitem{10}

Fluri, T. P., and T. W. Von Backström. "Performance analysis of the power conversion unit of a solar chimney power plant." Solar Energy 82, no. 11 (2008): 999-1008.


\bibitem{11}

Peter Vorobieff, Andrea Mammoli, Nima Fathi, and Vakhtang Putkaradze. "Free-standing inflatable solar chimney: experiment and theory." Bulletin of the American Physical Society 59 (2014).

\bibitem{12}

Nima Fathi, Peter Vorobieff, Seyed Sobhan Aleyasin. "V\&V Exercise for a Solar Tower Power Plant." ASME Verification and Validation Symposium (2014).


%https://cstools.asme.org/csconnect/FileUpload.cfm?View=yes&ID=44167


\bibitem{13}
Zhou, Xinping, Jiakuan Yang, Bo Xiao, Guoxiang Hou, and Fang Xing. "Analysis of chimney height for solar chimney power plant." Applied Thermal Engineering 29, no. 1 (2009): 178-185.

\bibitem{14}
Koonsrisuk, Atit, and Tawit Chitsomboon. "Theoretical turbine power yield in solar chimney power plants." In Thermal Issues in Emerging Technologies Theory and Applications (ThETA), 2010 3rd International Conference on, pp. 339-346. IEEE, 2010.

\bibitem{15}
Xu, Guoliang, Tingzhen Ming, Yuan Pan, Fanlong Meng, and Cheng Zhou. "Numerical analysis on the performance of solar chimney power plant system." Energy Conversion and Management 52, no. 2 (2011): 876-883.


\bibitem{16}
Tayebi, Tahar, and Mahfoud Djezzar. "Numerical Analysis of Flows in a Solar Chimney Power Plant with a Curved Junction." International Journal of Energy Science (2013).


\bibitem{kc13} Koonsrisuk, Atit, and Tawit Chitsomboon.
"Mathematical modeling of solar chimney power plants."
Energy 51 (2013): 314-322.




\bibitem{a}

Homayoon Daneshyar. "One-dimensional compressible flow." Oxford, Pergamon Press, Ltd., (1976).

\bibitem{b}

Iain G Currie. "Fundamental mechanics of fluids". McGraw-Hill, Inc., (1993).




\end{thebibliography}
\end{document}